\documentclass[12pt,epsf]{iopart}
\usepackage{graphicx}
\usepackage{amssymb,latexsym}
\usepackage{amsfonts}
\usepackage{amssymb}
\usepackage{cancel}
\usepackage{color}
\input{colordvi.tex}

\newcommand{\beqa}{\begin{eqnarray}}
\newcommand{\eeqa}{\end{eqnarray}}
\newcommand{\p}{\phi}

\def\Mpl{M_{\rm pl}}

\newcommand{\bef}{\begin{figure}}  \newcommand{\eef}{\end{figure}}
\newcommand{\bec}{\begin{center}}  \newcommand{\eec}{\end{center}}

\newcommand {\ga} {\ {\raise-.5ex\hbox{$\buildrel>\over\sim$}}\ }
\newcommand {\la} {\ {\raise-.5ex\hbox{$\buildrel<\over\sim$}}\ }

\def\MNRAS#1#2#3{Mon. Not. R. Astron. Soc. {\bf #1}, #2 (19#3)}

\begin{document}
%\draft

\title{Runaway Domain Wall and Space-time Varying $\alpha$}

\author{Takeshi Chiba}
\address{
%\affiliation{
Department of Physics, \\
College of Humanities and Sciences, \\
Nihon University, \\
Tokyo 156-8550, Japan}
\author{Masahide Yamaguchi}%
\address{
%\affiliation{
Department of Physics, 
Tokyo Institute of Technology, Tokyo 152-8551, Japan}

\date{\today}

\pacs{98.80.Cq ; 04.80.Cc }

\begin{abstract}
Recently spatial as well as temporal variations of the fine structure 
constant $\alpha$ have been reported. We show that a "runaway domain wall", 
which arises for the scalar field potential without minima, can account 
for such variations simultaneously. The time variation is induced by 
a runaway potential and the spatial variation is induced by the 
formation of a domain wall. 
The model is consistent with the current cosmological data and 
can be tested by the future experiments to test 
the equivalence principle. 
\end{abstract}

\maketitle

\section{Introduction}

String theory is the most promising approach to unify all the fundamental 
forces in nature. It is believed that in string theory all the 
coupling constants and parameters (except the string tension) in nature are 
derived quantities and are determined by the vacuum expectation values of 
the dilaton and moduli. However, only few mechanisms (see for example, \cite{douglas}) 
are known how and when to fix the dilaton/moduli. 
On the other hand, we know that the Universe is expanding. Then it is 
no wonder to imagine the possibility of the variation of the 
constants of nature during the evolution of the Universe. 

In fact, it is argued that the effective potentials of dilaton or moduli 
induced by nonperturbative effects may exhibit runaway structure; 
they asymptote zero for the weak coupling limit where dilaton becomes minus 
infinity or internal radius becomes infinity and symmetries are restored 
in the limit \cite{dine85,witten00}. Thus it is expected that as these 
fields vary, the natural ``constants'' may change in time 
and moreover the violation of the weak equivalence principle may be induced 
\cite{witten00,dp,dpv,dd,ckyy}. 

Hence, any detection or nondetection of such variations at various cosmological epochs 
could provide useful information about the physics beyond the standard model. 
In this respect, the recent claims of the detection of the time variation 
\cite{webb98,webb00,murphy03} as well as the spatial variation \cite{webb10} of the 
fine structure constant $\alpha$ may hint towards new physics. 

Narrow lines in quasar spectra are produced by absorption of radiation in 
intervening clouds of gas, many of which are enriched with heavy elements. 
Because quasar spectra contain doublet 
absorption lines at a number of redshifts, it is possible to check for 
time variation in $\alpha$ simply by looking for changes in the doublet 
separation of alkaline-type ions with one outer electron as a function of 
redshift. 

Webb et al. \cite{webb98} introduced a new technique (called
many-multiplet (MM) method) that compares the absorption wavelengths of
magnesium and iron atoms in the same absorbing cloud, which is far more
sensitive than the alkaline-doublet method.  From the latest analysis of
Keck/HIRES (High Resolution Echelle Spectrometer) 143 absorption systems
for $0.2<z<3.7$, they found that $\alpha$ was {\it smaller} in the past
\cite{murphy03}:
\beqa
\frac{\Delta \alpha}{\alpha} = (-0.543\pm 0.116)\times 10^{-5}.
\label{alpha:time}
\eeqa
Moreover, recently, Webb et al. analyzed a dataset from the ESO Very
Large Telescope (VLT) and found the opposite trend: $\alpha$ was {\it
larger} in the past \cite{webb10}.  Combined with the Keck samples, they
claimed the {\it spatial} variation of $\alpha$ \cite{webb10}:
\beqa
\frac{\Delta\alpha}{\alpha}=(1.10\pm0.25)\times 10^{-6} (r/{\rm Glyr})\cos\theta,
\label{alpha:space}
\eeqa
where $r$ is the look-back time $r=ct(z)$ and $\theta$ is the angle
between the direction of the measurement and the axis of best-fit
dipole.
  
However, concerning the claimed time variation of $\alpha$, similar
observations from VLT/UVES (Ultraviolet and Visual Echelle Spectrograph)
have not been able to duplicate these results
\cite{chand,chand2,levshakov04,molaro07,chand3}.  It is to be noted,
however, that the analysis by Srianand et al. \cite{chand} may suffer
from several flaws.  For example, the uncertainty in wavelength
calibration in \cite{chand} may not be consistent with the error in
$\Delta\alpha/\alpha$ \cite{levshakov05}.  According to the analysis of
the fundamental noise limitation \cite{levshakov05}, the systematic
errors in \cite{chand} may be several times underestimated. Recent
detailed re-analysis of Srianand et al. and Chand et al.  confirms these
concern: flawed parameter estimation methods in a $\chi^2$ minimization
analysis \cite{murphy071} (see however, \cite{srianand07}) and
systematic errors in the UVES wavelength calibration \cite{murphy07}.

Overall, although the claims of the detection of the variations are not
confirmed by independent methods (however see \cite{kanekar10}), the claims are also not disputed
seriously.  We regard the claims of the spatial/temporal variations
currently refuse to deny or confirm.  In this paper, we take the claims
of the spatial/temporal variations of $\alpha$ seriously and attempt to
explain the data qualitatively.

One major difficulty in explaining both temporal
(Eq. (\ref{alpha:time})) and spatial (Eq. (\ref{alpha:space})) 
variations is that the spatial variation across the horizon scale
($\Delta\alpha/\alpha\simeq 10^{-5}$ at $r\sim 14{\rm Glyr}$) is as
large as the time variation during the Hubble time.  If the time
variation is induced by a scalar field \cite{dz,ck,op,ag,cnp}, such a
cosmologically time evolving scalar field is very light with its mass
being comparable to the Hubble parameter and hence its relative 
fluctuation is very small. So it is almost impossible to explain
simultaneously both temporal and spatial variations by the light scalar
field and its fluctuation.  One possible solution to this problem is to
consider nonlinear objects like topological defects \cite{zko} or giant
voids \cite{inoue}. In this paper, we consider the former possibility
since the scalar field is dispersive and does not trace the matter
density perturbation much.\footnote{It is possible to induce the spatial variations by the environmental dependence \cite{op,mota}, but it is difficult to explain the variation of 
dipole type like Eq. (\ref{alpha:space}). } 
Ref. \cite{opu} considered a domain wall to
explain the spatial variations only but did not consider the time
variations. We point out a certain type of domain walls can be utilized
to explain not only the time variations of $\alpha$ but also 
the spatial variations of $\alpha$  of the same
order of magnitude as the time variations.

The paper is organized as follows: In Sec. 2, we detail our model and
then in Sec. 3 we study several constraints on the model parameters.
Sec. 4 is devoted to summary.

\section{Runaway Domain Wall and Varying $\alpha$}

For definiteness, we consider the theory described by the following action
\beqa
S=\int d^4x\sqrt{-g}\left[\frac{{\Mpl}^2}{2}R-
\frac12(\nabla \phi)^2-V(\phi)-
\frac{1}{16\pi \alpha_0} B(\p)F_{\mu\nu}F^{\mu\nu}
\right]+S_m.
\label{action}
\eeqa
Here $\Mpl=1/\sqrt{8\pi G}=2.4\times 10^{18}{\rm GeV}$ is the reduced
Planck mass, $\phi$ is the real symmetry breaking field,
$\alpha_0$ is the bare fine structure constant and $S_m$ denotes the
action of other matter (relativistic/non-relativistic particles and dark
energy).  We consider the following scalar field potential of runaway
type:
\beqa
V(\phi)=\frac{M^{2p+4}}{\left(\p^2+\sigma^2\right)^p}. 
\eeqa
Even if the potential has no minimum, the discrete symmetry
$\phi\leftrightarrow -\phi$ can be broken dynamically, which results in
the formation of a domain wall.  Such defects are dubbed "vacuumless"
defects in \cite{cv}, but we prefer to call them "runaway" defects since
they arise from the potential of runaway type.

\begin{figure}
\includegraphics[width=13cm]{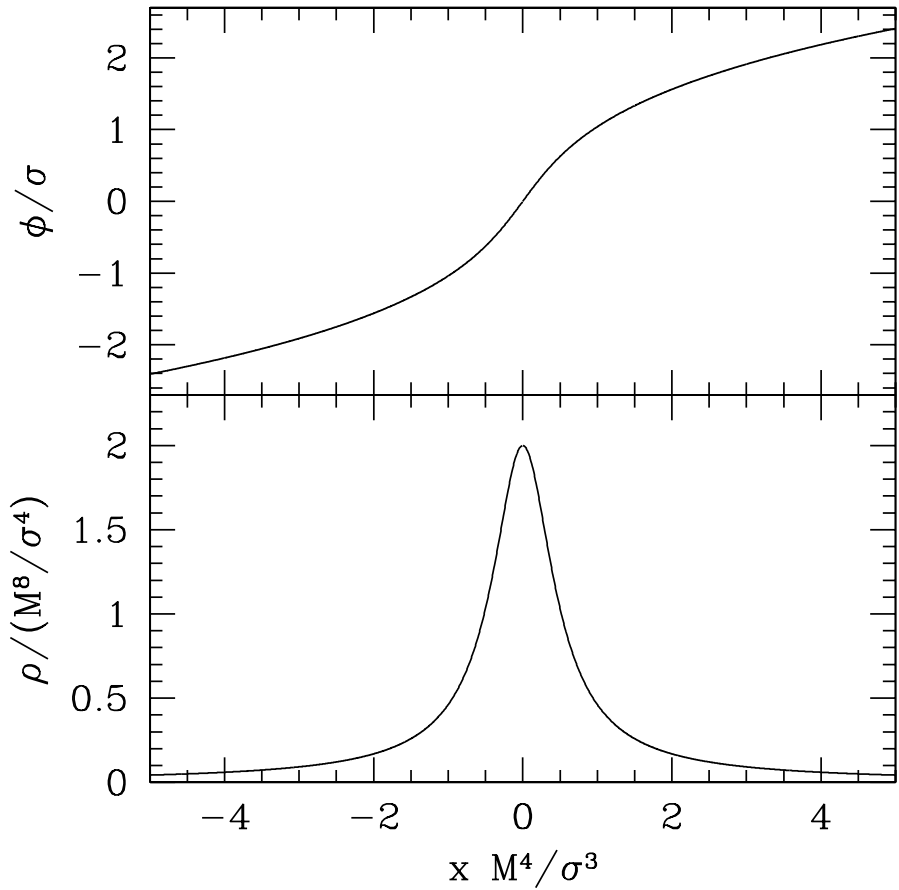}
\caption{$\p$ (upper) and the energy density (lower) of 
a runaway domain wall for $p=2$ in a flat space.   
}
\label{fig1}
\end{figure}

For $\p\gg \sigma$, from the balance between the kinetic energy and the
potential energy, one finds $\p \simeq (M^{p+2}x)^{1/(p+1)}$ and the
energy density is proportional to $x^{-2p/(p+1)}$, where $x$ is the
distance from the wall \cite{cv}. We assume $p>1$ so that the tension of
the wall is finite. Then, the width of the wall is estimated as $\delta
\simeq \sigma^{p+1}/M^{p+2}$ and the tension of the wall is given by
\beqa
\mu\simeq \frac{M^{2P+4}}{\sigma^{2p}}\delta\simeq \frac{M^{p+2}}{\sigma^{p-1}}.
\eeqa 
The profile of such a runaway domain wall solution is shown in
Fig. \ref{fig1}.  In the cosmological situation, we replace $x$ with the
Hubble distance $H^{-1}$ so that
\beqa
\p \propto a^{3/2(p+1)}
\label{tracker}
\eeqa 
during the matter dominated era in accord with the tracker solution for
$V\propto \p^{-2p}$ \cite{pr}. This scaling solution is useful to
account for the cosmological time variation of $\alpha$; in the opposite
case of $\p<\sigma$, the scalar field near the local maximum exhibits
thawing behavior \cite{chiba} and moves very slowly, and hence it is
difficult to explain the cosmological time variation of $\alpha$.

The coupling function $B(\p)$ in front of the electromagnetic kinetic term induces the spatio-temporal variations of $\alpha$ since $\alpha(\p)=\alpha_0/B(\p)$.   We consider the following coupling function :
\beqa
B(\p)=e^{-\xi\p/\Mpl}. 
\eeqa 
Since the effective $\alpha$ for small $\xi$ is given by 
\beqa
\alpha(\p)\simeq \alpha_0\left(1+\xi\frac{\p}{\Mpl}\right),
\label{alphaphi}
\eeqa
the time variation in either side of the wall is given by
\beqa
\frac{\dot\alpha}{\alpha_0}=\xi\frac{\dot\p}{\Mpl}=\pm \frac{3}{2(p+1)}\xi\frac{|\p|}{\Mpl}H,
\label{time}
\eeqa
where we have used Eq. (\ref{tracker}), and the spatial variation across the wall is given by
\beqa
\frac{\Delta\alpha}{\alpha_0}=\xi\frac{\Delta\p}{\Mpl}=
2\xi\frac{\p}{\Mpl}.
\label{space}
\eeqa
Thus, the opposite time variation of $\alpha$ between the Keck
(increasing $\alpha$) and the VLT (decreasing $\alpha$) as well as the
spatial variation of the same order of magnitude as the time variation
during the Hubble time ($\Delta\alpha/\alpha_0\sim
|\dot\alpha|/\alpha_0H^{-1}$) are accommodated in this model.

For definiteness, we shall choose $p=2$ henceforth so that $\p\propto a^{1/2}$. 
Then the measured $\alpha$ by the QSO absorption lines (\cite{webb98,webb00,murphy03,levshakov04,molaro07,levshakov05,murphy071}) are fitted 
as a function of $a^{1/2}$ as  
$(\alpha(a)-\alpha_0)/\alpha_0\simeq -6.2\times 10^{-6}(1-a^{1/2})$ and hence 
\beqa
\frac{\dot\alpha}{\alpha_0}\simeq 3.1\times 10^{-6}a^{1/2}H.
\label{datafit}
\eeqa
Using this fit in Eq. (\ref{time}) at $z\simeq 2$, we find that 
\beqa
\xi\frac{\p}{\Mpl}\simeq 3\times 10^{-6}. 
\label{xiphi}
\eeqa
Putting this into Eq. (\ref{space}), it implies that 
\beqa
\frac{\Delta\alpha}{\alpha_0}\simeq 7\times 10^{-6},
\eeqa
which explains naturally the largeness of the spatial variation.

Before closing this section, we would like to comment on the formation
and dynamics of domain walls. For $|\phi| \ll \sigma$, the (tree-level)
potential can be expanded as
\beqa
  V(\phi) \simeq V_0 - \frac12 m_{\phi}^2 \phi^2 + \frac14 \lambda \phi^4 +
            \cdots,
\eeqa
where $V_0 = M^8/\sigma^4,\,m_\phi^2 = 4 M^8/\sigma^6$, and $\lambda = 12
M^8/\sigma^8$. By taking into account the finite temperature effects,
the effective potential around the origin reads
\beqa
  V_{\rm T}(\phi) = V(\phi) + \xi^2 \frac{T^4}{\Mpl^2} \phi^2
                     + \frac{\lambda}{8}T^2 \phi^2 + \cdots,
\eeqa
Thus, the phase transition occurs around the critical temperature $T_c
\sim \frac{2}{\sqrt{3}} \sigma$ so that domain walls are formed. After
some relaxation period,\footnote{The friction force due to the thermal
plasma is estimated as $m_\phi^2 T^2 v$ ($v$ : wall velocity) and is
always subdominant in comparison to a force per unit area $\sim
\sigma/t$ coming from the curvature $\sim t$. Thus, the friction effects
on the domain wall dynamics due to the thermal plasma are
negligible. This is simply because the field $\phi$ consisting domain
walls only weakly interacts with the thermal plasma.} domain walls
evolve according to the linear scaling solution \cite{wall}, in which
typical scale of the domain wall is comparable to the Hubble scale and
its energy density is roughly given by
\beqa
  \rho_{\rm wall} \sim \mu H \propto \frac{1}{t}. 
\eeqa
%
%Here $\mu$ is the wall tension, that is, the energy per unit area and
%given by
%
%\beqa
%  \mu \sim \frac{M^8}{\sigma^4} \delta \sim \frac{M^4}{\sigma},
%\eeqa
%
%where $\delta \sim \sigma^3/M^4$ is the core width.

\section{Experimental Constraints}

Let us now discuss the observational and experimental constraints on the parameters. \\

\paragraph*{Sachs-Wolfe Effect:} 

A domain wall induces the temperature anisotropy by the Sachs Wolfe
effect \cite{zko}.  The gravitational potential due to the wall at the
horizon scale is $2\pi G\mu H_0^{-1}\simeq (1/4)\Mpl^{-2}\mu H_0^{-1}$
which induces the temperature anisotropy via the Sachs-Wolfe effect. The
requirement that this should be less than $10^{-5}$ gives
\beqa
M < 30 {\rm GeV} \left(\frac{\sigma}{10^{15}\rm GeV}\right)^{1/4}.
\label{bound}
\eeqa
The present energy density of a runaway domain wall within the horizon
scale is estimated as $\mu H_0^{-2}/(4\pi H_0^{-3}/3)$, which should be
much less than the critical density $3\Mpl^2H_0^2$. This also gives a
similar bound as Eq. (\ref{bound}).\\

\paragraph*{The Violation of the Weak Equivalence Principle:} 

Since the effective mass is very light: $\sqrt{V''}\simeq M^4/\p^3 <
10^{-41}{\rm GeV}(M/10 {\rm GeV})^4(\sigma/10^{15}{\rm GeV})^{-3}$ for
$\p> \sigma$, the scalar $\p$ mediates a long-range force via the
coupling to nucleons, leading to the violation of the weak equivalence
principle \cite{dz,ck}.  The modification of the nucleon mass follows
from the electromagnetic corrections.  To leading order in $\alpha$
these corrections are given by \cite{gl},
\beqa
&&\delta m_p=B_p\delta\alpha/\alpha_0=0.63{\rm MeV}\delta\alpha/\alpha_0,\\
&&\delta m_n=B_n\delta\alpha/\alpha_0=-0.13{\rm MeV}\delta\alpha/\alpha_0, 
\eeqa
where $m_p$ and $m_n$ are the proton and the neutron masses, and $B_p
\simeq 0.63$~MeV and $B_n \simeq -0.13$~GeV are the Born terms for the
proton and the neutron, respectively.\footnote{There also exists
corrections to the binding energy of nucleon \cite{gl,dz}, but the
effects do not change the result much. }  Hence, from
Eq. (\ref{alphaphi}), the exchange of $\p$ induces a composition
dependent long-range force.  A test body of mass $m$ experiences the
acceleration induced by the $\p$-exchange force \cite{dz,ck}
\beqa
a_{\p}=\frac{\xi^2}{4\pi \Mpl^2 m r^2}\left(N_p^EB_p+N_n^EB_n\right)\left(N_pB_p+N_nB_n\right),
\eeqa
where $N_{p,n}^E(N_{p,n})$ are numbers of protons and neutrons in the
Earth (the test body), in addition to the usual Newtonian acceleration
due to the Earth mass $M_E$: $a_g=M_E/8\pi\Mpl^2r^2$. The difference in
accelerations between the two test bodies in
E\"{o}tv\"{o}s-Dicke-Braginsky type experiments is parametrized by the
E\"{o}tv\"{o}s ratio: $\eta = 2{|a_1 - a_2|}/{|a_1 + a_2|}$, where $a_1$
and $a_2$ are the accelerations of two bodies.  Here we assume that the
test bodies have almost equal masses $m_1 \simeq m_2$, which implies
$N_{p,1}+N_{n,1} \simeq N_{p,2}+N_{n,2}$.  In the present case, $\eta$
is estimated as
\beqa
\eta\simeq \frac{\Delta a_{\p}}{a_g}\simeq 2\frac{\xi^2}{\overline{m}^2}
\left(\frac{N_p^EB_p+N_n^EB_n}{N_p^E+N_n^E}\right)
\left(\frac{\Delta N_p\,B_p+\Delta N_n\,B_n}{N_{p,1}+N_{n,1}}\right),
\eeqa
where we have used $m_i \simeq (N_{n,i} + N_{p,i})\overline{m}$ and $M_E
\simeq (N^E_n + N^E_p)\overline{m} $ with $\overline{m}$ being the
atomic mass unit ($\simeq 0.931$ GeV) and $\Delta N_p \equiv N_{p,1} -
N_{p,2},~\Delta N_n \equiv N_{n,1} - N_{n,2}$.  Adopting the typical
values for $N_{p,n}^E$ and $\Delta N_{p,n}$ \cite{dz,ck}, we find
\beqa
\eta\simeq 3\times 10^{-14}\left(\frac{\xi}{10^{-3}}\right)^2.
\label{eta:xi}
\eeqa
This should be smaller than the current experimental bounds
$\eta<2\times 10^{-13}$ \cite{wep}, which gives
\beqa
\xi < 2.6\times 10^{-3}.
\label{xi}
\eeqa

\begin{figure}
\includegraphics[width=13cm]{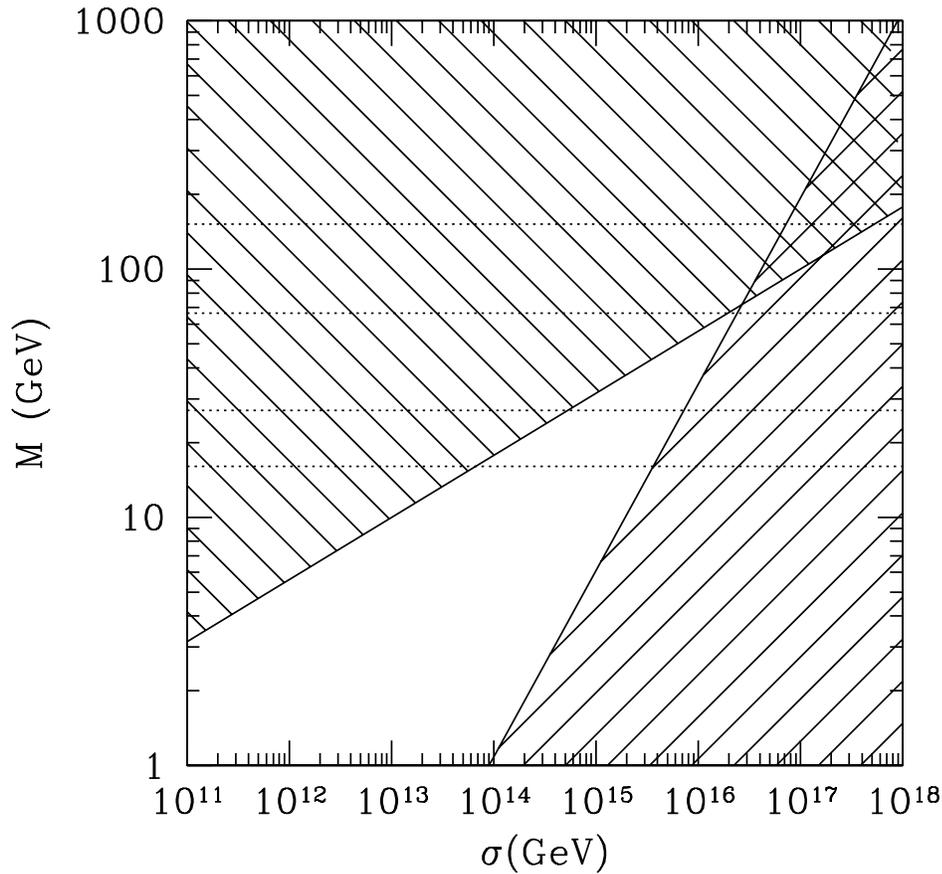}
\caption{Allowed parameter space. Upper region is excluded due to the Sachs-Wolfe effect (or large density parameter) Eq. (\ref{bound}); 
lower region is excluded because of $\p<\sigma$ and 
the absence of the scaling solution, Eq. (\ref{phisigma}). 
Dotted lines explain the QSO data (Eq. (\ref{mxi})) with $\xi=2\times 10^{-3},10^{-3},3\times 10^{-4},10^{-4}$ from bottom to top.    
}
\label{fig3}
\end{figure}

\paragraph*{Allowed Parameter Region:} 

The present value of $\p$ is estimated as
\beqa
\p_0\simeq (M^4H_0^{-1})^{1/3}.
\label{phi0}
\eeqa 
From Eq. (\ref{xiphi}) and Eq. (\ref{phi0}), 
we obtain 
\beqa
M\simeq 30 {\rm GeV}\left(\frac{\xi}{10^{-3}}\right)^{-3/4}.
\label{mxi}
\eeqa 
Moreover, the requirement of $\phi>\sigma$ to account for the time variation of $\alpha$, 
from Eq. (\ref{phi0}), leads to 
\beqa
M>6 {\rm GeV} \left(\frac{\sigma}{10^{15}\rm GeV}\right)^{3/4}.
\label{phisigma}
\eeqa

For example, for $\sigma\simeq 10^{15}{\rm GeV}$, $M\simeq 30 {\rm GeV}$
and $\xi\simeq 10^{-3}$ satisfy Eqs. (\ref{bound}), (\ref{mxi}), and
(\ref{phisigma}).  This gives $\p_0\simeq 8\times 10^{15}{\rm GeV}$.  In
Fig. \ref{fig3}, we show the allowed range of the parameters $M$ and
$\sigma$ together with the relation Eq. (\ref{mxi}).\\

\paragraph*{Window for Future Experiments:} 

From Eq. (\ref{bound}) and Eq. (\ref{phisigma}), upper bounds on $M$ and
$\sigma$ are found:
\beqa
M<70 {\rm GeV}~~~~~ {\rm and}~~~~~\sigma<2\times 10^{16}{\rm GeV},
\eeqa
which imply a lower bound on $\xi$ from
Eq. (\ref{mxi}):
\beqa
3\times 10^{-4}<\xi<2.6\times 10^{-3}.
\eeqa
This in turn provides a  window for $\eta$ from Eq. (\ref{eta:xi}):
\beqa
3\times 10^{-15}<\eta<2\times 10^{-13}.
\eeqa
Therefore, orders of magnitude improvements of the experimental limits
on the weak equivalence principle by proposed experiments (such as
MICROSCOPE \cite{microscope}, SR-POEM \cite{srpoem}, Galileo Galilei
\cite{gg} and STEP \cite{step}) could lead to the detection of the
violation of the weak equivalence principle induced by the scalar
exchange force or refute this model. These experiments (in particular,
MICROSCOPE launched in 2012) can test the violation of the equivalence
principle better than $\eta=10^{-15}$. Therefore, the model can be
tested within a few years by these gravitational experiments.

%\subsection{Further Example}

%We also entertain the possibility to present a model based on a domain wall with time dependent %tension \cite{gucci} where the time dependence is induced by the runaway dilaton:  
%
%\beq
%S=\int d^4x\sqrt{-g}\left[\frac12 R-\frac12(\nabla \varphi)^2-V_0e^{-\beta \varphi}-
%\frac12(\nabla \p)^2-\frac{\lambda}{4}(\p^2-\varphi^2)^2-\frac{1}{16\pi \alpha_0} B(\p)F_{\mu\nu}F^{\mu\nu}
%\right]+S_m.
%\label{action2}
%\eeq
%
%Here  $\varphi$ is the dilaton $S_m$ denotes the action of
%other matter (non-relativistic particles and dark energy). We use the units of $\Mpl=1$. 
% Spatio-temporal variations of $\alpha$ are 
%induced (mainly) by a domain wall with a time dependent tension: Its formation and dynamics 
%is described by $\p$ with the vacuum expectation value at $\p=\pm \varphi$ 
%and the time dependence of the tension is given by the dilaton $\varphi$. 

\section{Summary}

Motivated by possible detections of spatial and temporal variations of
$\alpha$, we have proposed a model based on a domain wall of runaway
type. We have found that it is possible to construct a model to explain
both variations simultaneously.  We have studied the cosmological
constraints on the model and found that the model can be made consistent
with the current cosmological data and can be falsified by the future
experiments to test the equivalence principle.  We note that a model is
not limited to a runaway potential, but we can construct a model with
local minima so that the vacuum expectation value is determined by a
runaway dilaton \cite{gucci}.

We have focused on $\alpha$ in this paper since our prime purpose was to
provide a existence proof of a model.  However, unless forbidden by
symmetry, the direct couplings of $\p$ to fermionic matter should exist,
which result in the violation of the equivalence principle and in
spatio-temporal variations of the proton-to-electron mass ratio. There
are some indications of a non-zero value of a spatial variation of it
\cite{lev10}. It would be interesting to study the consequences of such
matter couplings further.

%%%%%%%%%%%%%%%%%%%%%%%%%%%%%%%%%%%%%%%%%%%%%%%%%%%%%%%%%%%%%%%%%%%%%%
\ack
%\section*{Acknowledgments}
%%%%%%%%%%%%%%%%%%%%%%%%%%%%%%%%%%%%%%%%%%%%%%%%%%%%%%%%%%%%%%%%%%%%%%

This work was supported in part by a Grant-in-Aid for Scientific
Research from JSPS (No.\,20540280(TC) and No.\,21740187(MY)) 
and in part by Nihon University. 

%%%%%%%%%%%%%%%%%%%%%%%%%%%%%%%%%%%%%%%%%%%%%%

%%%%%%%%%%%%%%%%%%%%%%%%%%%%%%%%%%%%%%%%%%%%%%%%%%%
%%%%%%%%% references %%%%%%%%%%%%%%%%%%%%%%%%%%%%%%
\section*{References}
%%%%%%%%% references %%%%%%%%%%%%%%%%%%%%%%%%%%%%%%

\end{document}